\documentclass{aa520}

\usepackage[dvips]{graphicx}
\usepackage{epsfig}
\usepackage{txfonts}

\begin{document}
  
\title{An explanation for the kHz-QPO twin peaks separation in slow
and fast rotators.}

\author{J\'er\^ome P\'etri \inst{1}}

\offprints{J. P\'etri}

\institute {Max-Planck-Institut f\"ur Kernphysik, Saupfercheckweg 1,
    69117 Heidelberg, Germany.}

\date{Received / Accepted}

\titlerunning{An explanation for the kHz-QPO twin peaks separation}
\authorrunning{J. P\'etri}

\maketitle

\begin{abstract}
  
  In this Letter we further explore the idea, suggested previously by
  Klu{\'z}niak and collaborators, that the high frequency QPOs may be
  explained as a resonant coupling between the neutron star spin and
  two epicyclic modes of accretion disk oscillations. We confirm
  result of Lee, Abramowicz and Klu{\'z}niak (\cite{Lee2004}) that the
  strongest response occurs when the frequency difference of the two
  modes equals either the spin frequency (for ``slow rotators'') or half
  of it (for ``fast rotators''). New points discussed in this Letter
  are: (1) We suggest that the coupling is gravitational, and due to a
  non-axially symmetric structure of the rotating neutron star. (2) We
  found that two excited modes may be both connected to vertical
  oscillations of the disk, and that strong gravity is not needed to
  excite the modes.
  
\end{abstract}

\section{INTRODUCTION}

To date, quasi-periodic oscillations (QPOs) have been observed in
about twenty Low Mass X-ray Binaries (LMXBs) sources containing an
accreting neutron star. Among these systems, the high-frequency QPOs
(kHz-QPOs) which mainly show up by pairs, denoted by frequencies
$\nu_1$ and $\nu_2>\nu_1$, possess strong similarities in their
frequencies, ranging from~300~Hz to about~1300~Hz, as well as in their
shape~(see van der Klis~\cite{vanderKlis2000} for a review).  For slow
rotators (i.e. with rotation rate $\nu_* \approx 300$ Hz), the
frequency difference between the two peaks is around $\Delta \nu
\approx \nu_*$ and $\nu_2 \approx 3\,\nu_1/2$ whereas for fast rotators
(i.e. $\nu_*\approx 600$ Hz), this difference is around $\Delta \nu
\approx \nu_*/2$, (van der Klis~\cite{vanderKlis2004}).  For black
hole candidates the 3:2 ratio was first noticed by Abramowicz and
Klu{\'z}niak (\cite{Abramowicz2001}) who also recognized and stressed
its importance.  Now the 3:2 ratio of black hole QPOs frequencies is
well established (McClintock and Remillard, \cite{MacClintock2003}).

Many attempts have been made to explain this phenomenology. The
relativistic precession model introduced by Stella \& Vietri
(\cite{Stella1998}, \cite{Stella1999}) makes use of the motion of a
single particle in the Kerr-spacetime.  However, the peak separation
is not naturally deduced from their model. More promisingly,
Abramowicz \& Klu{\'z}niak~(\cite{Abramowicz2001}) introduced a
resonance between orbital and epicyclic motion that can account for
the 3:2 ratio around Kerr black holes leading to an estimate of their
mass and spin.  Klu{\'z}niak, Abramowicz, Kato et
al.~(\cite{Kluzniak2004a}) showed that the twin kHz-QPOs is explained
by a non linear resonance in the epicyclic motion of the accretion
disk.  Rebusco~(\cite{Rebusco2004}) developed the analytical treatment
of these oscillations. Bursa, Abramowicz, Karas \& Klu{\'
  z}niak~(\cite{Bursa2004}) suggested a gravitational lens effect
exerting a modulation of the flux intensity induced by the vertical
oscillations of the disk while simultaneously oscillating radially.
More recently, T{\" o}r{\" o}k, Abramowicz, Klu{\' z}niak \&
Stuchl{\'{\i}}k~(\cite{Torok2005}) applied this resonance to determine
the spin of some microquasars.

It was recognized (Klu{\'z}niak et al.~\cite{Kluzniak2004a}, Klu{\'
  z}niak, Abramowicz \& Lee~\cite{Kluzniak2004b}) that because in the
accreting millisecond pulsar SAX J1808.4-3658 the difference in
frequencies of the double peaked QPOs is clearly equal to half of the
pulsar spin (Wijnands, van der Klis, Homan et
al.~\cite{Wijnands2003}), the epicyclic resonance must be excited by a
coupling of the accretion disk oscillation modes to the neutron star
spin. Numerical simulations by Lee, Abramowicz \& Klu{\' z}niak
(\cite{Lee2004}) modelled the coupling by an unspecified external
forcing of the disk (with periodicity equal to that of the spin) and
found that resonant response occurs when the difference between
frequencies of the two modes equals to one-half of the spin frequency
(as observed in SAX J1808.4-3658 and other ``fast rotators''), and
when it equals to the spin frequency (as observed in ``slow rotators''
like XTE J1807-294).
  
In this Letter we further explore these ideas by showing that the
desired coupling may be provided by gravity of a sufficiently
non-axially symmetric neutron star. (Discussion of a physical
plausibility of non-axially symmetric neutron stars is beyond the
scope of this Letter).  In such a case, the accretion disk will
experience a gravitational field with dipolar, quadrupolar and
octupolar moments (m=1,2,3) that vary periodically in time.

\section{THE MODEL}
\label{sec:Model}

In this section, we describe the main features of the model, starting
with a simple treatment of the accretion disk, assumed to be made of
non interacting single particles orbiting in the equatorial plane of
the star. We thus neglect the hydrodynamical aspect of the disk like
pressure. Particles evolve in a perfectly spherically symmetric
gravitational potential until, at the origin of time $t=0$, a rotating
asymmetric part is added to the stellar gravitational field. This
perturbation is issued from an inhomogeneity in the neutron star
interior, for instance, as explained in the following subsection.

\subsection{Distorted stellar gravitational field}

We assume that the stellar interior is inhomogeneous and anisotropic.
In some regions inside the star, clumps of matter generate locally a
stronger or weaker gravitational potential than the average, depending
on their density. In order to compute analytically such kind of
gravitational field, we idealize this situation by assuming that the
star is made of an homogeneous and isotropic matter everywhere (with
total mass~$M_*$ and angular speed~$\Omega_*$).  To this perfect
spherically symmetric geometry, we add a small mass point, the
perturber having a mass~$M_\mathrm{p}\ll M_*$ located inside the star
at a position~$\vec{R}_\mathrm{p} = (r_\mathrm{p}, \varphi_\mathrm{p}
= \Omega_*\,t, z_\mathrm{p})$ with $r_\mathrm{p}^2 + z_\mathrm{p}^2
\le R_*^2$. We use cylindrical coordinates denoted by~$\vec{R}=(r,
\varphi, z)$.  The origin of the coordinate system coincides with the
location of the neutron star. A finite size inhomogeneity can then be
thought as a linear superposition of such point masses.  The total
gravitational potential induced by this idealized rotating
inhomogeneous star is~:
\begin{equation}
  \label{eq:PotBoiteux}
  \Phi(r,\varphi,z,t) = - G\,M_* / || \vec{R} || - 
  G\,M_\mathrm{p} / || \vec{R} - \vec{R}_\mathrm{p} ||
\end{equation}
where the first term in the right hand side corresponds to the
unperturbed spherically symmetric gravitational potential whereas the
second term is induced by the small point like inhomogeneity.  Using
the cylindrical frame of reference, the gravitational potential
reads~:
\begin{eqnarray}
  \label{eq:PotBoiteuxCyl}
  \Phi(r,\varphi,z,t) & = &
  - G\,M_* / \sqrt{r^2+z^2} - 
  G\,M_\mathrm{p} / \sqrt{r^2+z^2} \, \times \nonumber \\
  & & \sum_{m=0}^{+\infty} b_m^{1/2}\left( \sqrt{r_\mathrm{p}^2+z_\mathrm{p}^2} / \sqrt{r^2+z^2} \right)
  \cos(m\,\psi) 
\end{eqnarray}
where the azimuth in the corotating frame is~$\psi = \varphi -
\Omega_* \, t$ and the Laplace coefficients~$b_m^{n}(x)$ of
celestial mechanics are~:
\begin{eqnarray}
  \label{eq:CoeffLaplace}
  b_m^{n}(x) & = & \frac{2-\delta_m^0}{2\,\pi} \, \int_0^{2\,\pi} 
  \frac{ \cos(m\,\psi)}{(1 + x^2 - 2\,x\,\cos\psi)^n } \, d\psi
\end{eqnarray}
where $\delta_m^0$ is the Kronecker symbol. The total linear response
of the disk is then the sum of each perturbation corresponding to one
particular mode~$m$. Because the perturber is inside the star and the
disk never reaches the stellar surface, $x<1$ and thus the Laplace
coefficients~$b_m^{1/2}(x)$ never diverge. Moreover, because of the
term $\cos(m\,\psi)$ in the integrand Eq.~(\ref{eq:CoeffLaplace}), the
value of the Laplace coefficients decreases rapidly with the azimuthal
number~$m$. As a result, only the low azimuthal modes will influence
significantly the evolution of the disk. Keeping only the few first
terms in the expansion is sufficient to achieve reasonable accuracy.

\subsection{Equation of motion for a test particle}

All particles evolve in the gravitational field imposed by the neutron
star.  Their equation of motion reads
\begin{eqnarray}
  \label{eq:EquationMvtRad}
  \ddot{r} - r \, \dot{\varphi}^2 & = & g_\mathrm{r} + \delta g_\mathrm{r} \\
  \label{eq:EquationMvtAzi}
  2 \, \dot{r} \, \dot{\varphi} + r \, \ddot{\varphi} & = & \delta g_\varphi \\
  \label{eq:EquationMvtVer}
  \ddot{z} & = & g_\mathrm{z} + \delta g_\mathrm{z}
\end{eqnarray}
The gravitation field of the star~$M_\mathrm{*}$ is denoted
by~$\vec{g}$ whereas that of the perturber~$M_\mathrm{p}$ is denoted
by~$\delta\vec{g}$ and the dot means time derivative $d/dt$. We are
only interested in the vertical motion experienced by the test
particles in response to the vertical perturbed gravitational field.
We therefore neglect the radial and azimuthal perturbations, $\delta
g_\mathrm{r} = \delta g_\varphi = 0$.  We keep only the vertical
component, $\delta g_\mathrm{z}\ne0$.  According to this
simplification, the second equation then integrates immediately. It
states the conservation of the angular momentum of the particle $L =
m\,r^2\,\dot{\varphi} = \mathrm{const}$. This is the obvious integral
of motion for this problem.

Perturbing Eq.~(\ref{eq:EquationMvtVer}) and developing to first order
in the perturbation around the equilibrium Keplerian orbit defined by
$(r = r_0, \varphi = \Omega_\mathrm{k}\,t, z_0=0)$, the vertical
motion reads~:
\begin{eqnarray}
\label{eq:Oscillation}
  \ddot{z} + \left ( \Omega_\mathrm{k}^2 + \Omega_\mathrm{p}^2 
  \, \sum_{m=0}^{+\infty} b_m^{3/2\,(0)} \, 
  \cos \, ( m(\Omega_\mathrm{k}-\Omega_*) \, t) \right ) \, z & = & 
  \nonumber \\
  \Omega_\mathrm{k}^2 \, z_\mathrm{p} \, \sum_{m=0}^{+\infty} 
  b_m^{3/2\,(0)} \, \cos \, ( m \, (\Omega_\mathrm{k}-\Omega_*) \, t) 
\end{eqnarray}
where the Laplace coefficient are evaluated at the point~$(r_0,z_0)$,
$b_m^{3/2\,(0)} = b_m^{3/2} \left( \sqrt{r_\mathrm{p}^2 +
    z_\mathrm{p}^2}/\sqrt{r_0^2+z_0^2} \right)$, $\Omega_\mathrm{k} =
(G\,M_*/r_0^3)^{1/2}$ is the Keplerian orbital frequency and
$\Omega_\mathrm{p} = (G\,M_\mathrm{p}/r_0^3)^{1/2}$. We recognize a
Hill equation (periodic variation of the eigenfrequency of the system
on the left hand side) with a periodic driving force (on the right
hand side).

\subsection{Resonance conditions}

Eq.~(\ref{eq:Oscillation}) describes an harmonic oscillator with
periodically varying eigenfrequency which is also excited by a driven
force.  It is well known that some resonances will therefore occurs in
this system. Namely, we expect three kind of resonances corresponding
to~:
\begin{itemize}
\item a {\it corotation resonance} at the radius where the angular
  velocity of the test particle equals the rotation speed of the star.
  This is only possible for prograde motion. The resonance condition
  is $\Omega_\mathrm{k} = \Omega_* $;
\item a {\it driven resonance} at the radius where the vertical
  epicyclic frequency equals the frequency of each mode of the
  gravitational potential as seen in the locally corotating frame. The
  resonance condition is $m \, | \Omega_* - \Omega_\mathrm{k} | =
  \kappa_\mathrm{z}$;
\item a {\it parametric resonance} related to the time-varying
  vertical epicyclic frequency, (Hill equation). The rotation of the
  star induces a sinusoidally variation of the vertical epicyclic
  frequency leading to the well known Mathieu's equation for a given
  azimuthal mode $m$. The resonance condition are~:
  \begin{equation}
    \label{eq:ResPara}
    m \, |\Omega_* - \Omega_\mathrm{k}| = 2 \, \frac{\kappa_\mathrm{z}}{n}
  \end{equation}
\end{itemize}
where $n\ge 1$ is a natural integer.  Note that the driven resonance
is a special case of the the parametric resonance for~$n=2$. However,
their growth rate differ by the timescale of the amplitude
magnification.  Driving causes a linear growth in time while
parametric resonance causes an exponential growth. We also rewrite the
vertical epicyclic frequency as~$\kappa_\mathrm{z}$ instead of
$\Omega_\mathrm{k}$ in order to apply the results to a more general
case which could include magnetic field or general relativistic
effects. Indeed, in Sec.\ref{sec:Results} we will apply the
aforementioned resonance criteria also to the Kerr spacetime for which
the degeneracy between orbital and vertical epicyclic frequency is
lifted ($\kappa_\mathrm{z} \ne \Omega_\mathrm{k}$).

\section{RESULTS}
\label{sec:Results}

\subsection{Newtonian disk}
From Eq.~(\ref{eq:ResPara}), we can find the radius where each of this
resonance will occur.  Beginning with the Newtonian potential, it is
well known that the angular velocity and the vertical epicyclic
frequencies for a single particle are equal so that~$\Omega_\mathrm{k}
= \kappa_\mathrm{z}$.  This conclusion remains true for a thin
accretion disk having~$c_\mathrm{s} / r \, \Omega_\mathrm{k} \ll 1$
where $c_\mathrm{s}$ is the sound speed.  Distinguishing between the
two signs of the absolute value, we get for the parametric resonance
condition Eq.~(\ref{eq:ResPara}) the following orbital rotation rate~:
\begin{eqnarray}
  \frac{\Omega_\mathrm{k}}{\Omega_*} & = & \frac{m}{m \pm 2/n}
\end{eqnarray}
As a consequence, the resonances are all located in the frequency
range~$\Omega_\mathrm{k}\in[\Omega_*/3, 3\,\Omega_*]$.  In
Table~\ref{tab:ResPara}, we indicate the results for a slow as well as
for a fast rotator (respectively $\nu_*=300$~Hz and $\nu_*=600$~Hz).
Numerical applications are given for a spinning neutron star, showing
the first three modes~$m$ and the first two integers~$n=1,2$.
\begin{table}[h]
    \caption{Value of the orbital frequencies at the parametric
      resonance for the first three modes~$m$ and with $n=1,2$, in the
      case of a Newtonian gravitational potential.  The results are
      given for a~$1.4\,M_\odot$ neutron star rotating respectively
      at~$300$ and~$600$~Hz. The value on the left of the symbol~/
      corresponds to the absolute value sign taken to be~$-$ and on the
      right to be~$+$.}
  \label{tab:ResPara}
  \begin{center}
    \begin{tabular}{c c c | c c}
      \hline
      \hline
      Mode~$m$ & \multicolumn{4}{c}{Orbital frequency $\nu(r,a_*)$ (Hz)} \\
      \hline
      & \multicolumn{2}{c|}{$\nu_*=600$~Hz} & \multicolumn{2}{c}{$\nu_*=300$~Hz} \\
      & $n=1$ & $n=2$ & $n=1$ & $n=2$ \\
      \hline
      \hline
      1 & $-$600 / 200 & ---- / 300 & $-$300 / 100 & --- / 150 \\
      2 &   ---- / 300 & 1200 / 400 &    --- / 150 & 600 / 200 \\ 
      3 &   1800 / 360 &  900 / 450 &    900 / 180 & 450 / 225 \\
      \hline
    \end{tabular}
  \end{center}
\end{table}
The pair of highest orbital frequencies for the $\nu_* = 300$~Hz
spinning neutron star are $\nu_1 = 600$~Hz and $\nu_2 = 900$~Hz. The
twin peak separation frequency is then~$\Delta\nu = 300~\mathrm{Hz} =
\nu_*$.  The vertical motion induced by the parametric resonance at
that location will appear as a modulation in the luminosity of the
accretion disk. For the $\nu_* = 600$~Hz spinning neutron star, the
highest orbital frequencies are 1800~Hz and 1200~Hz. However, due to
the ISCO, the former one is not observed because it is located inside
the ISCO and therefore does not correspond to a stable orbit. Indeed,
remind that for a $1.4\,M_\odot$ neutron star, the maximal orbital
frequency in the Schwarzschild spacetime at the ISCO is
$\nu_\mathrm{isco} = 1571$~Hz. Therefore, the first two highest
observable frequencies are $\nu_1=900$~Hz and $\nu_2=1200$~Hz. This is
confirmed in the next subsection where the Newtonian field is replaced
by the Kerr geometry.  The peak separation frequency becomes
then~$\Delta\nu = 300$~Hz~$= \nu_*/2$. Thus the peak separation for
slow spinning neutron stars is $\Delta\nu = \nu_*$ whereas for fast
spinning neutron star it becomes $\Delta\nu = \nu_*/2$. This
segregation between slow and fast rotating neutron stars is well
observed in several accreting systems~(van der Klis
\cite{vanderKlis2004}). A similar reason for the dichotomy between
fast and slow rotators (location of the resonance radius), was
previously suggested by Lee, Abramowicz \& Klu{\' z}niak
(\cite{Lee2004}).

\subsection{General relativistic disk}

When the inner edge of the accretion disk reaches values of a few
gravitational radii, the general relativistic effects become
important. The degeneracy between the frequencies~$\Omega_\mathrm{k}$
and~$\kappa_\mathrm{z}$ is lifted and will depend on the angular
momentum of the star.  The characteristic orbital and vertical
epicyclic frequencies in the accretion disk around a Kerr black hole
(or a rotating neutron star) are the orbital angular velocity,
$\Omega(r,a_*) = G\,M_*/(r^{3/2} + R_\mathrm{g}^{3/2} \, a_*)$ where
$R_\mathrm{g} = G\,M_*/c^2$ is the gravitational radius and the
vertical epicyclic frequency,~$\kappa_\mathrm{z}(r,a_*) =
\Omega(r,a_*)\,\sqrt{1 -4\,(R_\mathrm{g}/r)^{3/2} \, a_* + 3 \,
  (R_\mathrm{g} / r )^2 \, a_*^2 }$. The parameter~$a_*$ corresponds
to the angular momentum of the star, in geometrized units.  For a
neutron star of mass~$M_*$, moment of inertia~$I_*$ and rotating at
the angular velocity~$\Omega_*$, it is given by $a_* = c\, I_* \,
\Omega_*/ G\, M_*^2$.  The parametric resonance conditions
Eq.~(\ref{eq:ResPara}) splits into two cases, depending on the sign of
the absolute sign~:
\begin{eqnarray}
\label{eq:ResParaGR}
   \Omega(r,a_*) \pm \frac{2\,\kappa_\mathrm{z}(r,a_*)}{m\,n} & = & \Omega_* 
\end{eqnarray}
For a given angular momentum~$a_*$, Eq.~(\ref{eq:ResParaGR}) have to
solved for the radius~$r$. For a neutron star, we adopt the typical
parameters~:
\begin{itemize}
\item mass~$M_*=1.4\,M_\odot$~;
\item angular velocity~$\nu_*=\Omega_*/2\pi=300/600$~Hz~;
\item moment of inertia~$I_*=10^{38}\;~\rm{kg\,m^2}$~;
\end{itemize}
The angular momentum is then given by~$a_* = 5.79 \times10^{-5} \,
\Omega_*$.  Solving Eq.~(\ref{eq:ResParaGR}) for the radius and then
deducing the orbital frequency at this radius we get the results shown
in Table~\ref{tab:ResParaGRver}. For the above chosen stellar spin
rate we find~$a_*=0.1/0.2$ for the slow and fast rotator respectively.
Therefore the vertical epicyclic frequency remains close to the
orbital one~$\kappa_\mathrm{z} \approx \Omega_\mathrm{k}$. As a
consequence, for the vertical resonance, the Newtonian approximation
mentioned in the previous subsection remains valid. Indeed, for a slow
rotator, the two highest frequencies are~$\nu_1 = 599$~Hz and $\nu_2 =
898$~Hz, therefore $\Delta\nu = 299$~Hz which is still very close
to~$\nu_*$ whereas for a fast rotator, $\nu_1 = 899$~Hz and $\nu_2 =
1198$~Hz, therefore $\Delta\nu = 299$~Hz which is close to~$\nu_*/2$
(the orbit located within the ISCO has been discarded).
\begin{table}[h]
  \caption{Same as Table \ref{tab:ResPara} but in the general
    relativistic Kerr spacetime.}
  \label{tab:ResParaGRver}
  \begin{center}
    \begin{tabular}{c c c | c c}
      \hline
      \hline
      Mode~$m$ & \multicolumn{4}{c}{Orbital frequency $\nu(r,a_*)$ (Hz)} \\
      \hline
      & \multicolumn{2}{c|}{$\nu_*=600$~Hz} & \multicolumn{2}{c}{$\nu_*=300$~Hz} \\
      & $n=1$ & $n=2$ & $n=1$ & $n=2$ \\
      \hline
      \hline
      1 & ---- / 200 & ---- / 300 &  --- / 100 & --- / 150 \\
      2 & ---- / 300 & 1198 / 400 &  --- / 150 & 599 / 200 \\ 
      3 & 1790 / 360 &  899 / 450 &  898 / 180 & 450 / 225 \\
      \hline
    \end{tabular}
  \end{center}
\end{table}
Numerical simulations in which the disk was disturbed by an external
periodic field confirmed this point of view (Lee, Abramowicz \& Klu{\'
  z}niak~\cite{Lee2004}).  We emphasize the fact that these results
apply to a rotating asymmetric magnetic field with exactly the same
resonance conditions Eq.~(\ref{eq:ResPara}) provided that the flow is
not to far from its Keplerian motion, i.e. a weakly magnetized
accretion disk with high~$\beta$-plasma parameter.  Note also that
these results are quiet general and {\it independent of the spacetime
  geometry}. General relativity is not required to account for the 3:2
ratio.  This model therefore also encompasses the accreting white
dwarfs for which QPOs have been observed in the same 3:2 ratio and
show strong similarities with X-ray binaries, (Warner \&
Woudt~\cite{Warner2005}).

\section{CONCLUSION}
\label{sec:Conclusion}

In this Letter, the consequences of a weak rotating asymmetric
gravitational potential perturbation on the evolution of a thin
accretion disk initially in a stationary axisymmetric state have been
explored. For gravitational perturbation with multipolar components,
the response of the disk is the sum of individual modes as long as it
remains in the linear regime. The physical processes at hand does not
require any general relativistic effect. Indeed, the resonances behave
identically in the Newtonian as well as in the Kerr field. As a
consequence, the QPO phenomenology is unified in a same picture,
whatever the nature of the compact object. Indeed, observations in
accretion disks orbiting around white dwarfs, neutron stars or black
holes have shown a strong correlation between their low and high
frequencies QPOs (Mauche~\cite{Mauche2002}, Psaltis, Belloni \& van
der Klis~\cite{Psaltis1999}). The relation is found to be same
irrespective of the nature of the compact object. This strongly
supports the idea of one and same general mechanism at hand in this
accretion disks.  The twin peaks ratio around 3:2 for the kHz-QPOs is
naturally explained not only for black hole candidates or neutron
stars but also for white dwarfs as reported by Warner, Woudt \&
Pretorius~(\cite{Warner2003}).  Indeed, the presence or the absence of
a solid surface, a magnetic field or an event horizon play no relevant
role in the production of the X-ray
variability~(Wijnands~\cite{Wijnands2001}). The twin peak separation
being either $\Delta\nu = \nu_*$ for slow rotator ($\nu_* \approx
300$~Hz) or $\Delta\nu = \nu_*/2$ for fast rotator ($\nu_* \approx
600$~Hz) is also explained.  Indeed, it was previously argued by
Klu{\'z}niak, Lasota, Abramowicz \& Warner (\cite{Kluzniak2005}) that
the 3:2 QPOs recently observed in white dwarf sources by Warner \&
Woudt (\cite{Warner2005}) have the same nature as the strong gravity
3:2 QPOs observed in neutron star and black hole sources. All of them
are resonant accretion disk oscillations.  Differences are attributed
to different modes that are involved, to different mechanisms of
resonance excitation, and to different modulations of the X-ray flux.

To conclude, to date we know about 20~LMXBs containing a neutron star
and all of them show kHz-QPOs. These QPOs can be explained by a
mechanism similar to those exposed here. We need only to replace the
gravitational perturbation by a magnetic one as described in
P\'etri~(\cite{Petri2005}). However, in an accreting system in which
the neutron star is an oblique rotator, we expect a perturbation in
the magnetic field to the same order of magnitude than the unperturbed
one. Therefore, the linear analysis developed in this paper has to be
extended to oscillations having non negligible amplitude compared to
the stationary state. Nonlinear oscillations therefore arise naturally
in the magnetized accretion disk, leading to a shift in the resonance
criteria and accounting for a change in the peak separation in
relation with the accretion rate.

\begin{acknowledgements}
  I am grateful to the referee Marek A. Abramowicz for his valuable
  comments and remarks. This work was supported by a grant from the
  G.I.F., the German-Israeli Foundation for Scientific Research and
  Development.
\end{acknowledgements}

\end{document}